# Pressure-induced Topological Node-Line Semimetals in Alkaline-Earth Hexaborides $X$B$_6$ ($X$=Ca, Sr, Ba)

L.-Y. Gan,[1,2] R. Wang,[1,3] Y. J. Jin,[1] D. B. Ling,[4] J. Z. Zhao,[1] W. P. Xu,[1] J. F. Liu,[1,*] and H. Xu[1,†]

[1]*Department of Physics, South University of Science and Technology of China, Shenzhen 518055, China*

[2]*Key Laboratory of Advanced Technology of Materials (Ministry of Education), Superconductivity and New Energy R&D Center, Southwest Jiaotong University, Chengdu, 610031 Sichuan, China*

[3]*Institute for Structure and Function & Department of Physics, Chongqing University, 400030 Chongqing, China*

[4]*Department of Physics, Anhui University, Hefei 230601, China*

**Abstract**

Based on first-principles calculations, we reported that external pressure can induce topological phase transition in alkaline-earth hexaborides, $X$B$_6$ (X=Ca, Sr, Ba). It was revealed that $X$B$_6$ is transformed from trivial semiconductors to topological node-line semimetals under moderate pressures when spin-orbit coupling (SOC) is ignored. The band inversion between B $p_x$ ($p_z$) and $p_y$ orbitals at X point is responsible for the formation of node-line semimetals. Three node-line rings around X point are protected by the combination of the time-reversal and spatial inversion symmetries, and the "drumhead" surface bands are obtained in the interiors of the projected node-line rings. When SOC is included, tiny gaps (< 4.8 meV) open at the crossing lines, and the $X$B$_6$ becomes strong topological insulators with $Z_2$ indices (1;111). As the SOC-induced gap opening is negligible, our findings thus suggest ideal real systems for experimental exploration of the fundamental physics of topological node-line semimetals.

## 1. Introduction

The hexaboride compounds $X$B$_6$ (X = Ca, Sr, Ba, La, Ce, Sm, Eu, and Gd) have been extensively studied over the last few decades due to their various physical properties.[1-11] Particularly, broad interests have been attracted to explore the topological nature of rare-earth hexaborides due to their strong correlations of $f$ electrons.[7-11] A consensus has been reached on SmB$_6$, which is a topological Kondo insulator at low temperatures as a result of the hybridization between $d$ and $f$ orbitals of Sm ions.[12] Recently, efforts have been made to resolve the controversial issue in YbB$_6$: whether its electronic structure is nontrivially topological or not. Kang *et al.* suggested that YbB$_6$ is a trivial semiconductor because Yb has a fully filled 4$f$ shell, but becomes a $p$-$d$ overlap semimetal under pressure.[11] Despite the intensive interest in the topological nature of the rare-earth hexaborides, however, to date no attention has been paid to the topology of *alkaline-earth* hexaborides to the best of our knowledge, which crystallize in the same structure as that of SmB$_6$ and YbB$_6$ but exclude strong correlated $d$ or $f$ electrons.

Even if no symmetry is broken by smoothly varying material parameters, changes of the topological order and quantum phase transitions are still likely to occur.[13-15] Alkaline-earth hexaborides were long thought to be simple polar semiconductors.[16] La-doped CaB$_6$ exhibits an unexpected high-temperature ferromagnetism despite the absence of partly filled $d$ or $f$ orbitals.[17] To explain this observation, early theoretical studies suggested a small overlap in the vicinity of X point of the simple cubic Brillouin zone, indicating that the compounds are *semimetallic*.[1, 17, 18] However, later quasi-particle calculations of the single-particle excitation spectrum[3] and experimental techniques[4] revealed an energy gap between the valence and conduction bands at X point. These experimental and theoretical uncertainties have prevented a consensus on the nature of alkaline-earth hexaborides.

With the successful discovery of topological insulators, enormous attention has been paid to new types of quantum matter in condensed matter physics and materials science: topological semimetals.[19] They are classified in three categories: Dirac semimetal,[20-24] Weyl semimetal,[25-31] and node-line semimetal (NLSM).[32-38] Specifically, the third one is characterized by a fully closed line of the crossing between energy bands in the Brillion zone near the Fermi level ($E_F$).[19] The origin is band inversion, which leads to node-line loops protected by the coexistence of time-reversal and inversion symmetry when spin-orbit coupling (SOC) is ignored.[19, 32, 34, 39] The novel exotic properties are promising for high-temperature superconductivity.[40, 41]

In this work, based on first-principles calculations, we reported that alkaline-earth hexaborides $X$B$_6$ ($X$=Ca, Sr, Ba) are intrinsic trivial semiconductors, but become topologically nontrivial semimetals under moderate pressures due to the band inversion between the bonding and anti-bonding states at X point. As a result, alkaline-earth hexaborides are NLSMs with "drumhead" like surface bands when SOC is turned off. Particularly, SOC opened gaps at the crossing lines are negligible because the bonding and anti-bonding states at X point are dominated by B $p$ orbitals, rendering these compounds ideal systems for experimental exploration of the fundamental physics of topological node-line semimetals.

## 2. Computational details

First-principles calculations are performed using the Vienna Ab initio Simulation Package with Perdew-Burke-Ernzerhof generalized gradient approximation (GGA-PBE)[42-44] and the Heyd-Scuseria-Ernzerhof hybrid exchange correlation functional (HSE06).[45] A cutoff energy of 500 eV and a 11 × 11 × 11 $k$-mesh are used to optimize the structure until all residual forces remain below 0.01 eV/Å, while a Γ-centered 23 × 23 × 23 $k$-mesh is adopted for total energy and electronic structure calculations. The surface band structures are calculated in a tight-binding scheme based on the maximally localized Wannier functions, which are projected from the bulk Bloch wave functions.[46]

## 3. Results and discussion

In alkaline-earth hexaboride compounds $X$B$_6$, $X$ refers to Ca, Sr and Ba. Figure 1(a) shows their cubic structure, which belongs to $Pm\bar{3}m$ (No. 221) symmetry. It consists one $X$ atom in the corner and six B atoms in the interior of the cell to form a regular octahedron. It is obvious that CaB$_6$, SrB$_6$, and BaB$_6$ crystallize in the same structure and share similar electronic features: band structures and pressure-dependent band inversion, as shown in Fig. S1 (CaB$_6$ and SrB$_6$) in the supplementary materials and in Fig. 2 (BaB$_6$). Therefore, we present results of BaB$_6$ in detail here.

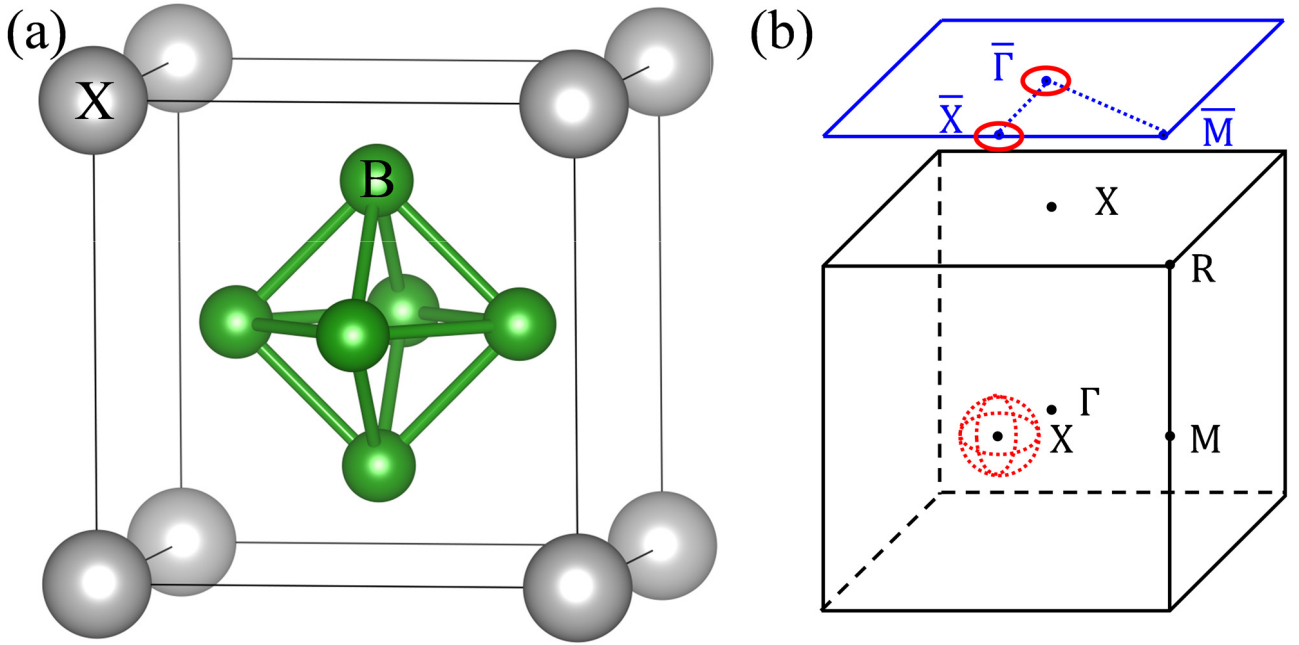

FIG. 1 (Color online). (a) Crystal structure of $X$B$_6$. (b) Bulk and projected (001) surface Brillouin zone. The dotted red circles show the scheme of the three node-line rings around the face center X point. They are perpendicular to each other and parallel to $k_x = 0$, $k_y = 0$, and $k_z = 0$ plane, respectively. The solid red circles show the schemes of projected "drumhead" states at $\bar{\Gamma}$ (0, 0) and $\bar{X}$ (0.5, 0) points.

Table I. Parity product of occupied states at the time-reversal invariant momenta points for $X$B$_6$: R, Γ, M (×3), and X (×3). HSE06, GGA-PBE, and GGA-PBE-SOC results of $Z_2$ indices, and $Z_2$ under a lattice strain with a = 0.90 $a_0$ calculated by HSE06 (indicated by HSE06$^p$).

| Functionals | R | Γ | X (×3) | M (×3) | $Z_2$ |
|---|---|---|---|---|---|
| HSE06 | - | + | - | + | (0:000) |
| HSE06$^p$ | - | + | + | + | (1:111) |
| GGA-PBE | - | + | + | + | (1:111) |
| GGA-PBE-SOC | - | + | + | + | (1:111) |

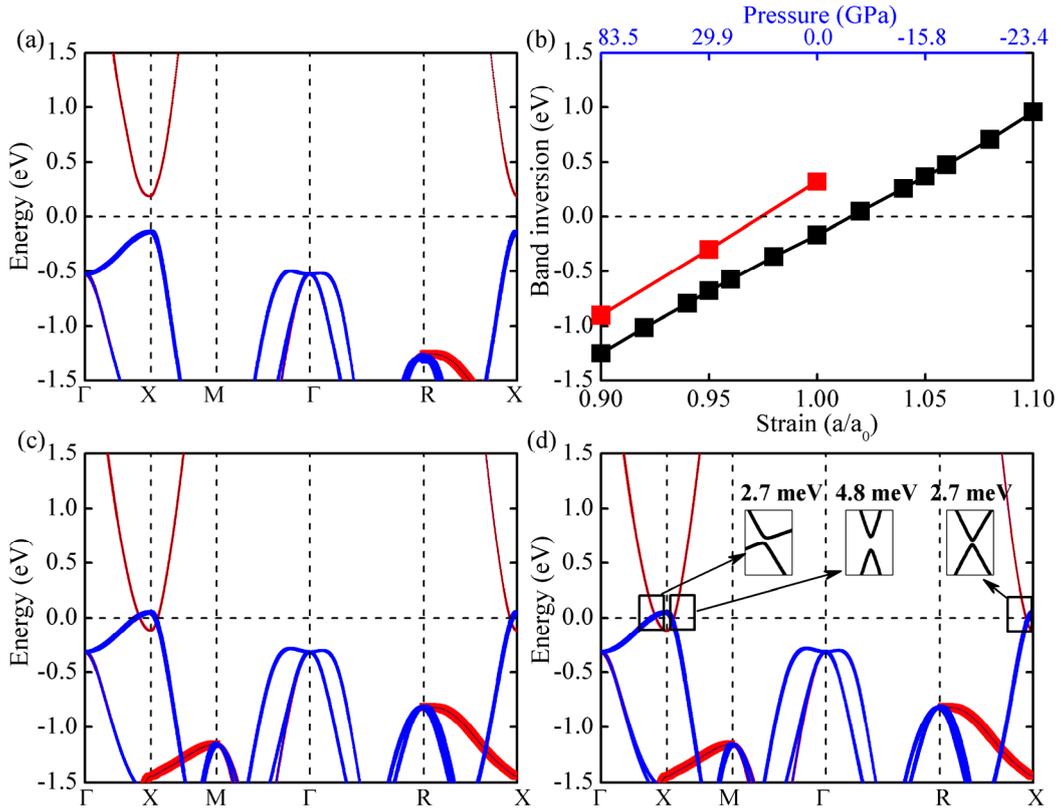

FIG. 2 (Color online) Projected band structures of BaB$_6$ from (a) HSE06, (c) GGA-PBE without SOC, and (d) GGA-PBE with SOC. The blue and red color show B $p_x$ ($p_z$) and $p_y$ orbitals, respectively. The insets in (d) show the SOC induced gaps. (b) Evolution of the band inversion between the bonding and anti-bonding bands at X point. Negative values indicate the existence of band inversion. Black: GGA-PBE. Red: HSE06.

Early theoretical studies of $X$B$_6$ ($X$ = Ca, Sr) suggested a small overlap in the vicinity of X point of the simple cubic Brillouin zone, indicating a semimetallic nature.[17,18] However, this description is challenged by later quasi-particle calculations of single-particle excitation spectrum[3] and experimental observation,[4] which indicated that $X$B$_6$ is semiconducting and the gap is giant compared to the SOC induced gaps in the semimetallic description. To ascertain this controversy, HSE06 calculations is employed to shed light on the electronic structure of BaB$_6$, as shown in Fig. 2(a). The system is a semiconductor with an obvious direct band gap (0.32 eV) at X point. Further band projection analysis suggests that the two bands are dominated by B $p_x$ ($p_z$) and $p_y$ orbital, respectively (see Fig. 2(a)). Table I lists the parity product of occupied states at the time-reversal invariant momenta points to investigate the topology of BaB$_6$. The topological $Z_2$ indices are (0:000). Therefore, BaB$_6$ is an intrinsic topological trivial semiconductor, consistent with previous experimental report.[4]

A recently report suggested quantum phase transitions occur in YbB$_6$ single crystal under pressure: from a topologically trivial semiconductor to a topologically nontrivial semimetal and then to a possible topologically nontrivial gapped state.[10] As alkaline-earth hexaborides crystallize in the same structure, we thus turn to whether or not similar transitions occur in alkaline-earth hexaborides under pressure. Figure 2(b) exhibits the pressure-dependent energy difference between the bonding (B $p_x$ ($p_z$)) and anti-bonding (B $p_y$) bands at X point. Negative values indicate that band inversion occurs. Both GGA-PBE and HSE06 results are shown. It can be seen that they actually exhibit a similar trend: band inversion increase as pressure increases. By HSE06 calculations, we find that band inversion happens when a ≤ ~0.98 a$_0$. Namely, an external pressure of ~10 GPa is capable to induce the inversion between the bonding and anti-bonding bands at X point and thus a possible phase transition in BaB$_6$. The GGA-PBE obtained band structure is displayed in Fig. 2(c). Except for the feature around X point, the band structures are quite similar to those predicted by HSE06, especially the orbital contribution of the energy bands (see Figs. 2(a) and 2(c)). As a matter of fact, the GGA-PBE obtained semimetallic description coincides with the results under pressure (~0.96 $a_0$) predicted by HSE06 due to its common underestimation of band gaps.[47] Specifically, such coincidence enables us to further discuss the properties induced by band inversion under pressure on the basis of the results of GGA-PBE.

Based on parity analysis about the eigenvalues of the Bloch wave function (see Table I), we find that the bonding and anti-bonding states at X point have opposite parities. Therefore, the band inversion transforms the system from a trivial semiconductor to a topologically nontrivial NLSM, confirming a phase transition, similar to that in YbB$_6$.[10, 11] Moreover, the topologically nontrivial NLSM is also confirmed by HSE06 calculations listed in Table I. Three crossing points are found slightly above or below $E_F$ along Γ-X, X-M, and R-X, respectively. As a result, three node-line loops around X point are obtained (see Fig. 1b) with the coexistence of time-reversal and inversion symmetry in the absence of SOC, similar to those in anti-perovskites.[34, 35] Additionally, though higher pressure results in larger band inversion, the conduction band minimum at X point does not yet touch any other bands under even higher pressure, i.e., a = 0.90 $a_0$ (P = 83.5 GPa). This is unlike the situation in YbB$_6$, in which Yb 4$d$ band ultimately mixes with Yb 4$f$ band under 50 GPa,[11] suggesting only one quantum phase transition in BaB$_6$ in such a wide range of pressures.

We further construct a minimal effective Hamiltonian that can describe three nearly circular-like node lines around X point. When the spin-orbit coupling is ignored, these node lines are protected by the coexistence of time reversal and spacial inversion symmetry.[36] Around X point, the two crossing bands are mainly from $p_x$ ($p_z$) and $p_y$ orbitals of six B atoms. A general two-band Hamiltonian is used to describe the two crossing bands,

$$H(\mathbf{k}) = g_0(\mathbf{k})\sigma_0 + g_x(\mathbf{k})\sigma_x + g_y(\mathbf{k})\sigma_y + g_z(\mathbf{k})\sigma_z$$

where $\mathbf{k}=(k_x,k_y,k_z)$, $g_{0,x,y,z}(\mathbf{k})$ are real functions. $\sigma_0$ is unit matrix and $\sigma_{x,y,z}$ are Pauli matrices acting in the space of the two crossing bands. The coexistance of time-reversal and inversion symmetry requires that $g_y = 0$, $g_0$ and $g_z$ being even function of $\mathbf{k}$, and $g_x$ being odd function of $\mathbf{k}$.[36] By considering the crystal symmetry at X point, we obtain the symmetry-allowed expressions for $g_0(\mathbf{k})$, $g_x(\mathbf{k})$, $g_z(\mathbf{k})$ up to the lowest order of $\mathbf{k}$,

$$g_0(\mathbf{k}) = M_0 - B_0(k_y^2 + k_z^2) - C_0 k_x^2,$$
$$g_x(\mathbf{k}) = A k_x k_y k_z,$$
$$g_z(\mathbf{k}) = M_z - B_z(k_y^2 + k_z^2) - C_z k_x^2.$$

The eigenvalues of this effective Hamiltonian are $E(\mathbf{k}) = g_0(\mathbf{k}) \pm \sqrt{g_x^2(\mathbf{k}) + g_z^2(\mathbf{k})}$. Node lines appear when $g_x(\mathbf{k}) = g_z(\mathbf{k}) = 0$. $g_z(\mathbf{k}) = 0$ requires $M_z B_z > 0$ and $M_z C_z > 0$, which is just the band inversion condition. At the same time, $g_x(\mathbf{k}) = 0$ confines the node lines in three mutually perpendicular planes (namely, $k_{x,y,z} = 0$ planes). Furthermore, the term $g_0(\mathbf{k})$ confirms the little energy dispersion of node lines.

In the presence of SOC, all the band crossing lines disappear, exhibiting tiny energy gaps, as shown by the insets in Fig. 2(d). The parities of the occupied states at eight time-reversal invariant momenta points are shown in Table I. The topological $Z_2$ indices are (1:111), indicating a strong topological insulator, similar to that found in inversion-symmetric $Cu_3NX$ crystals reported by Kim and coworkers.[35] However, because the two bands, where the band inversion occurs, are dominated by B $p$ orbitals, the SOC induced gaps are negligible (< 4.8 meV or 55 K) in liquid nitrogen condition. Breaking the symmetry gives a similar picture. Such small SOC opened gaps together with a not high pressure (~10 GPa) render these real compounds ideal systems for experimental exploration of the fundamental physics of NLSM.

The band inversion at X point under pressure and the three node lines in $BaB_6$ indicate the existence of topologically nontrivial surface states. In order to study the surface states a 400-unit-cells-thick (001) slab with the surface terminated by Ba atoms is employed. Because the SOC in the system is tiny enough to be ignored, the system is almost perfect NLSM, thus yielding nearly flat surface bands around $E_F$ in the enclosed region of the projected node-line rings. The Brillouin zone of the (001) surface is shown in Fig. 1b.

Two projection points are derived from bulk X point onto the two dimensional Brillouin zone, namely, $\bar{\Gamma}$ and $\bar{X}$, as seen in Fig. 1(b). The surface band structures are presented in Fig. 3. The nearly flat surface "drumhead" states are nestled inside the projected node-line rings around $\bar{\Gamma}$ and $\bar{X}$ points near $E_F$. Such nearly flat bands are proposed as a venue to achieve high-temperature superconductivity.[39, 40]

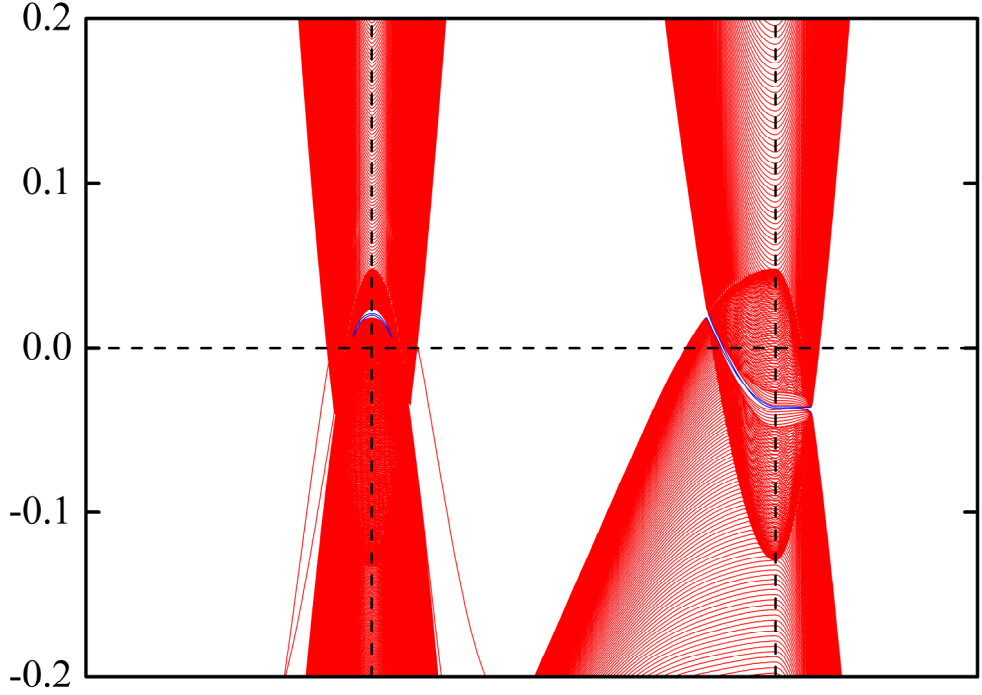

FIG. 3 (Color online). Surface states without SOC of the (001) surface of $BaB_6$. The "drumhead" states around $\bar{\Gamma}$ and $\bar{X}$ points are highlighted by the blue lines.

4. Conclusions

In summary, based on *ab initio* density functional theory calculations, topological NLSM states are proposed in alkaline-earth hexaborides $XB_6$ (X=Ca, Sr, Ba) under pressure. Three node-line rings are found around X point because of band inversion and protected by the coexistence of time-reversal and inversion symmetry. Due to the absence of strong correlated *d* and *f* elections, NLSM exists in $XB_6$ in a wide range of pressures. The "drumhead" surface bands are nestled inside the projected node-line rings around $\bar{\Gamma}$ and $\bar{X}$ points. In the presence of SOC, the system becomes a strong topological insulator with the topological $Z_2$ indices to be (1:111). Because band inversion occurs between B $p$ orbital-derived bonding and anti-bonding bands at X point, SOC induced gaps are negligible (< 4.8 meV). Our results not only suggest the existence of nontrivial topology in hexaboride compounds without strong electron correlations but provide ideal real systems for experimental explorations of fundamental physics in NLSMs.

**Acknowledgements**

This work is supported by the National Natural Science Foundation of China (NSFC, Grant Nos. 11674148, 11334003, 11504303, and 11404159), the Fundamental Research Funds for the Central Universities (SWJTU2682016ZDPY10), and the Basic Research Program of Science, Technology and Innovation Commission of Shenzhen Municipality (Grant No. JCYJ20160531190054083).

**Supporting Information**

Supplementary information (SI) available: band structures and pressure-dependent band inversion of $CaB_6$ and $SrB_6$.

*liujf@sustc.edu.cn

†xu.h@sustc.edu.cn

Reference

1. Hasegawa, A.; Yanase, A. *J. Phys. C* **1979,** 12, (24), 5431.
2. Massidda, S.; Monnier, R.; Stoll, E. *Eur. Phys. J. B* **2000,** 17, (4), 645-649.
3. Tromp, H. J.; van Gelderen, P.; Kelly, P. J.; Brocks, G.; Bobbert, P. A. *Phys. Rev. Lett.* **2001,** 87, (1), 016401.
4. Denlinger, J. D.; Clack, J. A.; Allen, J. W.; Gweon, G. H.; Poirier, D. M.; Olson, C. G.; Sarrao, J. L.; Bianchi, A. D.; Fisk, Z. *Phys. Rev. Lett.* **2002,** 89, (15), 157601.
5. Cho, B. K.; Rhyee, J.-S.; Oh, B. H.; Jung, M. H.; Kim, H. C.; Yoon, Y. K.; Kim, J. H.; Ekino, T. *Phys. Rev. B* **2004,** 69, (11), 113202.
6. Rhyee, J.-S.; Cho, B. K. *J. Appl. Phys.* **2004,** 95, (11), 6675-6677.
7. Kim, D. J.; Xia, J.; Fisk, Z. *Nat. Mater.* **2014,** 13, (5), 466-470.
8. Li, G.; Xiang, Z.; Yu, F.; Asaba, T.; Lawson, B.; Cai, P.; Tinsman, C.; Berkley, A.; Wolgast, S.; Eo, Y. S.; Kim, D.-J.; Kurdak, C.; Allen, J. W.; Sun, K.; Chen, X. H.; Wang, Y. Y.; Fisk, Z.; Li, L. *Science* **2014,** 346, (6214), 1208-1212.
9. Neupane, M.; Xu, S.-Y.; Alidoust, N.; Bian, G.; Kim, D. J.; Liu, C.; Belopolski, I.; Chang, T. R.; Jeng, H. T.; Durakiewicz, T.; Lin, H.; Bansil, A.; Fisk, Z.; Hasan, M. Z. *Phys. Rev. Lett.* **2015,** 114, (1), 016403.
10. Zhou, Y.; Kim, D.-J.; Rosa, P. F. S.; Wu, Q.; Guo, J.; Zhang, S.; Wang, Z.; Kang, D.; Yi, W.; Li, Y.; Li, X.; Liu, J.; Duan, P.; Zi, M.; Wei, X.; Jiang, Z.; Huang, Y.; Yang, Y.-f.; Fisk, Z.; Sun, L.; Zhao, Z. *Phys. Rev. B* **2015,** 92, (24), 241118.
11. Kang, C.-J.; Denlinger, J. D.; Allen, J. W.; Min, C.-H.; Reinert, F.; Kang, B. Y.; Cho, B. K.; Kang, J. S.; Shim, J. H.; Min, B. I. *Phys. Rev. Lett.* **2016,** 116, (11), 116401.
12. Dzero, M.; Sun, K.; Galitski, V.; Coleman, P. *Phys. Rev. Lett.* **2010,** 104, (10), 106408.
13. Hasan, M. Z.; Kane, C. L. *Rev. Mod. Phys.* **2010,** 82, (4), 3045-3067.
14. Qi, X.-L.; Zhang, S.-C. *Rev. Mod. Phys.* **2011,** 83, (4), 1057-1110.
15. Zhu, Z.; Cheng, Y.; Schwingenschlögl, U. *Phys. Rev. Lett.* **2012,** 108, (26), 266805.
16. Johnson, R. W.; Daane, A. H. *J. Chem. Phys.* **1963,** 38, (2), 425-432.
17. Zhitomirsky, M. E.; Rice, T. M.; Anisimov, V. I. *Nature* **1999,** 402, (6759), 251-253.
18. Rodriguez, C. O.; Weht, R.; Pickett, W. E. *Phys. Rev. Lett.* **2000,** 84, (17), 3903-3906.
19. Hongming, W.; Xi, D.; Zhong, F. *J. Phys.: Condens. Matter* **2016,** 28, (30), 303001.
20. Wang, Z.; Weng, H.; Wu, Q.; Dai, X.; Fang, Z. *Phys. Rev. B* **2013,** 88, (12), 125427.
21. Yang, B.-J.; Nagaosa, N. *Nat. Commun.* **2014,** 5.
22. Neupane, M.; Xu, S.-Y.; Sankar, R.; Alidoust, N.; Bian, G.; Liu, C.; Belopolski, I.; Chang, T.-R.; Jeng, H.-T.; Lin, H.; Bansil, A.; Chou, F.; Hasan, M. Z. *Nat. Commun.* **2014,** 5.
23. Borisenko, S.; Gibson, Q.; Evtushinsky, D.; Zabolotnyy, V.; Büchner, B.; Cava, R. J. *Phys. Rev. Lett.* **2014,** 113, (2), 027603.
24. Liu, Z. K.; Zhou, B.; Zhang, Y.; Wang, Z. J.; Weng, H. M.; Prabhakaran, D.; Mo, S.-K.; Shen, Z. X.; Fang, Z.; Dai, X.; Hussain, Z.; Chen, Y. L. *Science* **2014,** 343, (6173), 864-867.
25. Burkov, A. A.; Balents, L. *Phys. Rev. Lett.* **2011,** 107, (12), 127205.
26. Singh, B.; Sharma, A.; Lin, H.; Hasan, M. Z.; Prasad, R.; Bansil, A. *Phys. Rev. B* **2012,** 86, (11), 115208.
27. Huang, S.-M.; Xu, S.-Y.; Belopolski, I.; Lee, C.-C.; Chang, G.; Wang, B.; Alidoust, N.; Bian, G.; Neupane, M.; Zhang, C.; Jia, S.; Bansil, A.; Lin, H.; Hasan, M. Z. *Nat. Commun.* **2015,** 6.

28. Xu, S.-Y.; Belopolski, I.; Alidoust, N.; Neupane, M.; Bian, G.; Zhang, C.; Sankar, R.; Chang, G.; Yuan, Z.; Lee, C.-C.; Huang, S.-M.; Zheng, H.; Ma, J.; Sanchez, D. S.; Wang, B.; Bansil, A.; Chou, F.; Shibayev, P. P.; Lin, H.; Jia, S.; Hasan, M. Z. *Science* **2015,** 349, (6248), 613-617.
29. Weng, H.; Fang, C.; Fang, Z.; Bernevig, B. A.; Dai, X. *Phys. Rev. X* **2015,** 5, (1), 011029.
30. Lv, B. Q.; Weng, H. M.; Fu, B. B.; Wang, X. P.; Miao, H.; Ma, J.; Richard, P.; Huang, X. C.; Zhao, L. X.; Chen, G. F.; Fang, Z.; Dai, X.; Qian, T.; Ding, H. *Phys. Rev. X* **2015,** 5, (3), 031013.
31. Zheng, H.; Xu, S.-Y.; Bian, G.; Guo, C.; Chang, G.; Sanchez, D. S.; Belopolski, I.; Lee, C.-C.; Huang, S.-M.; Zhang, X.; Sankar, R.; Alidoust, N.; Chang, T.-R.; Wu, F.; Neupert, T.; Chou, F.; Jeng, H.-T.; Yao, N.; Bansil, A.; Jia, S.; Lin, H.; Hasan, M. Z. *ACS Nano* **2016,** 10, (1), 1378-1385.
32. Burkov, A. A.; Hook, M. D.; Balents, L. *Phys. Rev. B* **2011,** 84, (23), 235126.
33. Xie, L. S.; Schoop, L. M.; Seibel, E. M.; Gibson, Q. D.; Xie, W.; Cava, R. J. *APL Mater.* **2015,** 3, (8), 083602.
34. Yu, R.; Weng, H.; Fang, Z.; Dai, X.; Hu, X. *Phys. Rev. Lett.* **2015,** 115, (3), 036807.
35. Kim, Y.; Wieder, B. J.; Kane, C. L.; Rappe, A. M. *Phys. Rev. Lett.* **2015,** 115, (3), 036806.
36. Weng, H.; Liang, Y.; Xu, Q.; Yu, R.; Fang, Z.; Dai, X.; Kawazoe, Y. *Phys. Rev. B* **2015,** 92, (4), 045108.
37. Hu, J.; Tang, Z.; Liu, J.; Liu, X.; Zhu, Y.; Graf, D.; Myhro, K.; Tran, S.; Lau, C. N.; Wei, J.; Mao, Z. *Phys. Rev. Lett.* **2016,** 117, (1), 016602.
38. Li, R.; Ma, H.; Cheng, X.; Wang, S.; Li, D.; Zhang, Z.; Li, Y.; Chen, X.-Q. *Phys. Rev. Lett.* **2016,** 117, (9), 096401.
39. Heikkilä, T. T.; Volovik, G. E. *JETP Lett.* **2011,** 93, (2), 59-65.
40. Kopnin, N. B.; Heikkilä, T. T.; Volovik, G. E. *Phys. Rev. B* **2011,** 83, (22), 220503.
41. Volovik, G. E. *Phys. Scr.* **2015,** 2015, (T164), 014014.
42. Kresse, G.; Furthmüller, J. *Phys. Rev. B* **1996,** 54, (16), 11169-11186.
43. Kresse, G.; Furthmüller, J. *Comput. Mater. Sci.* **1996,** 6, (1), 15-50.
44. Perdew, J. P.; Burke, K.; Ernzerhof, M. *Phys. Rev. Lett.* **1996,** 77, (18), 3865-3868.
45. Heyd, J.; Scuseria, G. E.; Ernzerhof, M. *J. Chem. Phys.* **2006,** 124, (21), 219906.
46. Marzari, N.; Mostofi, A. A.; Yates, J. R.; Souza, I.; Vanderbilt, D. *Rev. Mod. Phys.* **2012,** 84, (4), 1419-1475.
47. Chan, M. K. Y.; Ceder, G. *Phys. Rev. Lett.* **2010,** 105, (19), 196403.